\documentclass[prl,twocolumn, showpacs,superscriptaddress]{revtex4}
\usepackage{graphicx,amsfonts,amsmath,amssymb}
\usepackage{times,txfonts}

\def\kbar{\protect\@kbar}
\def\@kbar{%
\relax \bgroup
\def\@tempa{\hbox{\raise.73\ht0
\hbox to0pt{\kern.25\wd0\vrule width.5\wd0
height.1pt depth.1pt\hss}\box0}}%
\mathchoice{\setbox0\hbox{$\displaystyle k$}\@tempa}%
{\setbox0\hbox{$\textstyle k$}\@tempa}%
{\setbox0\hbox{$\scriptstyle k$}\@tempa}%
{\setbox0\hbox{$\scriptscriptstyle k$}\@tempa}%
\egroup}

\begin{document}

\title{Quantum Accelerator Modes from the Farey Tree}
\author
{A. Buchleitner} \affiliation{ Max-Planck-Institut f\"{u}r Physik
Komplexer Systeme, D-01187 Dresden, Germany}
\author
{M. B. d'Arcy}\thanks{ Present address: Department of War Studies,
King's College London, London WC2R 2LS, UK} \affiliation{ Atomic
Physics Division, National Institute of Standards and Technology,
Gaithersburg, MD 20899-8424, USA}
\author{S. Fishman}
\affiliation{Physics Department, Technion, Haifa IL-32000, Israel}
\author{S. A. Gardiner}\thanks{
Present address: Department of Physics, Durham University, Durham
DH1 3LE, UK} \affiliation{JILA, University of Colorado and
National Institute of Standards and Technology, Boulder, CO
80309-0440, USA}
\author{I. Guarneri}
\affiliation{Dipartimento di Fisica e Matematica, Universit\`{a}
degli Studi dell'Insubria, I-22100 Como, Italy}
\affiliation{Instituto Nazionale per la Fisica della Materia,
Unit\`{a} di Milano, I-20133 Milano, Italy}
\affiliation{Instituto
Nazionale di Fisica Nucleare, Sezione di Pavia, I-27100 Pavia,
Italy}
\author{Z.-Y. Ma}
\affiliation{Department of Physics, University of Oxford, Oxford,
OX1 3PU, United Kingdom}
\author{L. Rebuzzini}
\affiliation{Dipartimento di Fisica e Matematica, Universit\`{a}
degli Studi dell'Insubria, I-22100 Como, Italy}
\author{G. S. Summy}
\affiliation{Department of Physics, Oklahoma State University, Stillwater,
OK 74078-3072, USA}

\date{\today}

\begin{abstract}
We show that mode-locking finds a purely \emph{quantum}
non-dissipative counterpart in atom-optical quantum accelerator
modes. These modes are formed by exposing cold atoms to periodic
kicks in the direction of the gravitational field. They are
anchored to generalized Arnol'd tongues, parameter regions where
driven nonlinear \emph{classical} systems exhibit mode-locking. A
hierarchy for the rational numbers known as the Farey Tree
provides an ordering of the Arnol'd tongues and hence of
experimentally observed accelerator modes.
\end{abstract}


\maketitle

Precise control of the state and time evolution of quantum systems
is of critical importance in many areas of physics.
Tailoring wave packets in Rydberg systems \cite{weinacht98},
producing single photons on demand \cite{brattke00}, creating
coherent superpositions of macroscopic persistent-current states
\cite{wal00}, and controlling the production of multiparticle
entanglement \cite{roos04}, are prominent examples of
``quantum state engineering''. Although almost
perfect control has been achieved over these systems, this can
rapidly lose efficiency when influenced by decoherence
or noise. Additionally, generic features of
strongly coupled quantum systems allow for novel and often
robust strategies of quantum control. In such cases,
studied in much detail in the area of quantum chaos, peculiar
eigenstates emerge which exhibit unexpected localization
properties and dynamics, and are remarkably inert with respect to
uncontrolled perturbations. Prominent examples are
nondispersive wave packets in periodically driven quantum systems
\cite{maeda04}, quantum resonances \cite{oskay2000,darcy2001b},
and stochastic web states \cite{fromholdnat,arrc04_1}. These
``strong coupling'' quantum control schemes rely on underlying
classical dynamics, which in general is mixed regular-chaotic \cite{abu02}.
For such a picture to be
meaningful it is in general necessary to approach the
semiclassical limit where the classical actions accumulated along
typical eigenmodes of the system are large compared to $\hbar$.
The quantum system can then ``resolve'' the
intricate phase space structure of classically mixed
regular-chaotic dynamics, and classical nonlinear stabilization
phenomena emerge on microscopic scales.

\begin{figure}[ht]
\centering
\includegraphics[width=8cm]{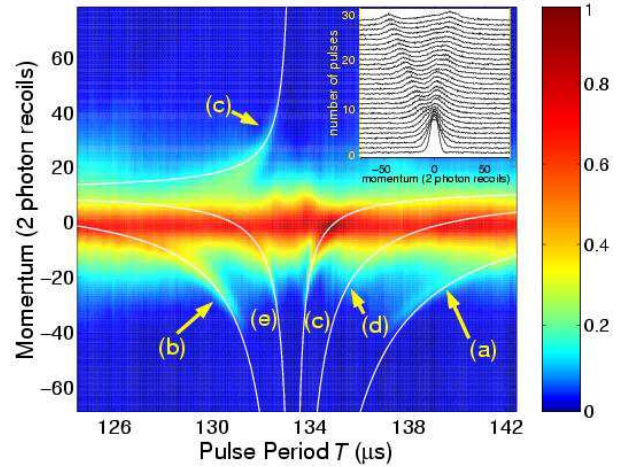}
\caption{Density plot of experimental atomic momentum
distributions (measured in a frame falling freely with $g$) after
$n=30$ pulses as $T$ is varied in the vicinity of
$T_{2}=133.3\mu$s, from $124.5 \mu$s to $142.5 \mu$s in steps of
$0.128 \mu$s \cite{schlunk2003b}. The labels (a), (b), and (c)
indicate $(2,1)$, $(3,1)$, and $(5,2)$ quantum accelerator modes,
as predicted by the corresponding Arnol'd tongues depicted in
Fig.\ \ref{Fig:simpletongues}. Labels (d) and (e) indicate $(7,3)$
and $(8,3)$ modes. White lines are the momenta predicted by Eq.\
(\ref{Eq:acceleration}). Inset shows the evolution of a typical
(in this case a $(1,0)$) QAM as a function of pulse number. The
color bar indicates the population scale.} \label{Fig:highorder}
\end{figure}

One of the most ubiquitous of such stabilization phenomena in
nonlinear classical dynamics is {\em mode-locking}. Eigenmodes of
a periodically driven dissipative system are locked in their
time-evolution onto the phase of an external drive through a
nonlinear resonance phenomenon. It occurs in applications ranging
from frequency-stabilized lasers \cite{sargent} to plasma
confinement in fusion reactors \cite{chirikov74}. An important
question is: is this necessarily strictly a
classical/semiclassical effect? We report that features of
mode-locking occur at the \emph{quantum level} even far from the
semiclassical limit. This was achieved using quantum accelerator
modes (QAMs) of cold atoms kicked by a pulsed standing wave of
light orientated along the Earth's gravitational field
\cite{schlunk2003b}. In this Letter we show that the modes
observed in these experiments can be classified according to a
number-theoretic construction known as the Farey Tree
\cite{farey}.

We commence by describing the experimental system which forms the
basis of our analysis \cite{schlunk2003b}. To create quantum
accelerator modes in the laboratory, laser cooled cesium atoms are
exposed to a sequence of equally spaced pulses from a standing
wave of light which is far detuned from the nearest atomic
transition. Due to the AC Stark effect, the atoms experience each
$\delta$-function like pulse as a sinusoidal potential (spatial
period $\lambda/2$) proportional to the intensity of the light. A
QAM is characterized by a momentum transfer, to a substantial
fraction of the atoms, which increases linearly with the number of
pulses. Figure \ref{Fig:highorder} shows the momentum of the
atomic ensemble as a function of pulse period $T$ with the
accelerator modes highlighted by the labelled curves
\cite{schlunk2003b}. The inset plots the momentum distribution of
a $(1,0)$ QAM as a function of the number of pulses. Note how the
width and amplitude of the accelerator mode (the peak which moves
to the left) remain relatively unchanged and the center momentum
increases linearly with pulse number. This can be contrasted with
the peak near zero momentum which broadens as pulses are applied.
Although the stability of the accelerator mode from pulse-to-pulse
is already somewhat reminiscent of mode locking behavior, there
are other reasons to think of the accelerator mode in this way.

To see this, we examine the atomic center-of-mass
dynamics using the one-dimensional $\delta$-kicked accelerator
Hamiltonian
$\hat{H} =
\hat{p}^2/2m + mg\hat{z} - \hbar \phi_d[ 1+\cos(G\hat{z})] \sum_n
\delta(t - nT)$ \cite{darcy2001a}. Here $\hat{z}$ is the vertical
position, $\hat{p}$ the momentum, $m$ the atomic mass, $g$ the
gravitational acceleration, $t$ the time, $G=4\pi/\lambda $
\cite{NoteRecoil}, and $\phi_d=U_0 t_p/2\hbar $, where $U_{0}$ is
the maximum AC Stark shift in the standing wave and $t_p$ is the
pulse duration. The special resonant values of $T$ in the vicinity
of which QAMs occur experimentally are $T_{\ell}=2\pi \ell
m/(\hbar G^2)=\ell \times 66.7\mu$s, with $\ell$ any non-negative
integer. Hence, $\epsilon=2\pi
\ell(T/T_{\ell}-1)$ is a small parameter.
Translating to a frame accelerating with $g$, to remove the linear
potential, and taking the limit
$\epsilon\to 0$
the quantum
dynamics of the kicked atoms can be modelled by the
classical map
\cite{FGR2002a}:
\begin{subequations}
\begin{align}
J_{n+1}=&J_{n}- K\sin(\theta_{n})  -
\mbox{sgn}(\epsilon)2\pi\Omega , \label{Eq:genericmap_a}
\\
\theta_{n+1}=&\theta_{n} +
\mbox{sgn}(\epsilon)J_{n+1}\;\;\mbox{mod}(2\pi),
\label{Eq:genericmap_b}
\end{align}
\label{Eq:genericmap}
\end{subequations}
where $\mbox{sgn}(\epsilon)$ is positive/negative if the pulse
interval $T$ is greater/smaller than $T_{\ell}$, and
\begin{subequations}
\begin{align}
\theta =& G z\;\;\mbox{mod}(2\pi)
\\
J_n=&I_n+\mbox{sgn}(\epsilon)[\pi
\ell+\beta\tau-2\pi\Omega(n+1/2)],
\end{align}
\label{Eq:Jvsn}
\end{subequations}
with $p/\hbar G=I/|\epsilon| + \beta$, $K=\phi_d|\epsilon|$,
$\Omega=gGT^2/2\pi$, $\tau=2\pi \ell T/T_{\ell}$.
Note that the deeply quantum mechanical character of the atomic dynamics
remains hidden in the parametrization of (\ref{Eq:genericmap},\ref{Eq:Jvsn})
through the quasimomentum $\beta$, $0\leq\beta <1$, since the limit
$\epsilon\to 0$ leading to the classical equations (\ref{Eq:genericmap})
leaves the finite value of $\hbar$ unaffected.
By Bloch theory, subspaces of different quasimomenta
are decoupled.

Mode-locking enters the theory of QAMs via Eq.\thinspace
(\ref{Eq:genericmap}) which also describes the deterministic
motion of a periodically kicked classical particle on a circle. In
this case $\theta_{n}$ and $J_{n}$ are the angle and angular
momentum just before the $n$th kick, $K$ is the kicking strength,
and $\Omega$ the unperturbed winding number. If the classical
particle is additionally subject to dissipative forces, the
accessible phase space shrinks, and Eq.\ (\ref{Eq:genericmap})
reduces (in the long-time limit) to the sine-circle map
\cite{schuster}: $\theta_{n+1}=\theta_{n}-K\sin
(\theta_{n})-2\pi\Omega$, a paradigm in the study of mode-locking.
If $K = 0$ and $\Omega$ is a rational number
$\mathfrak{m}/\mathfrak{p}$, any trajectory of the sine-circle map
returns to its initial value (modulo $2\pi$) after $\mathfrak{p}$
iterations. For $0<K<1$, mode-locking is observed; over a range of
$\Omega$ values around $\mathfrak{m}/\mathfrak{p}$ (the
mode-locking interval) a periodic trajectory with rational winding
number $\mathfrak{m}/\mathfrak{p}$ {\em persists}. This orbit
attracts all other orbits asymptotically in time, such that
finally all have this winding number. The widths of the
mode-locking intervals are exponentially small in $\mathfrak{p}$,
and increase with increasing $K$ up until $K=1$. The regions thus
formed in $(\Omega,K)$ parameter space, terminating at $K=0$,
$\Omega=\mathfrak{m}/\mathfrak{p}$, are known as {\em Arnol'd
tongues\/} \cite{jensen84}.

Using this formalism, it is now possible to analyze the dynamics
of the QAM in the $\epsilon \to 0$ limit. To begin we look for
stable periodic orbits in Eq.\thinspace (\ref{Eq:genericmap}) such
that, if $(J_{0},\theta_{0})$ is on an order $\mathfrak{p}$ orbit,
after $\mathfrak{p}$ pulses $J_{\mathfrak{p}}$
$\mbox{mod}(2\pi)=J_{0}$. If the orbit is stable, then each of the
$\mathfrak{p}$ points it is composed of is surrounded by a
nonlinear resonance island, set in a chaotic sea, where the motion
is predominantly regular; the motion in the island system
approximates that of the periodic orbit. If the orbit has winding
number $\mathfrak{m}/\mathfrak{p}$, then
$J_{\mathfrak{p}}=J_{0}+2\pi\mathfrak{m}$. Thus, from Eq.\
(\ref{Eq:Jvsn}), $I_{n}$ (and therefore $p_{n}$) grows linearly
with time. The islands travel in momentum, resulting in
acceleration. If a wavepacket is launched within an island
surrounding a stable periodic orbit, the acceleration of the
corresponding QAM obeys
\begin{equation}
p_{n} \simeq p_{0} + n\frac{2\pi}{|\epsilon|} \left(\Omega -
\frac{\mathfrak{m}}{\mathfrak{p}} \right)\hbar G\ ,
\label{Eq:acceleration}
\end{equation}
precisely as observed in the inset of Fig.~\ref{Fig:highorder}.
Hence,  we can identify QAMs with nonlinear resonance islands in the
classical phase space generated by (\ref{Eq:genericmap}), what is just another
manifestation of the general mode-locking phenomenon we are describing here.
These islands are robust structures, as
guaranteed by the Kolmogorov-Arnol'd-Moser (KAM) theorem
\cite{LL92}. A quantum wavepacket initially prepared in the island
travels with it, and decays only slowly by tunneling into the
chaotic surroundings. For sufficiently small $\epsilon$, this
tunneling is exponentially weak, resulting in a stable QAM. More
importantly, due to the KAM theorem the island itself is rather
inert with respect to perturbations of the Hamiltonian generating
the map Eq.\ (\ref{Eq:genericmap}). This robustness is inherited
by the QAMs and shields them against experimental noise
\cite{maeda04}.

\begin{figure}[tbp]
\centering
\includegraphics[width=8cm]{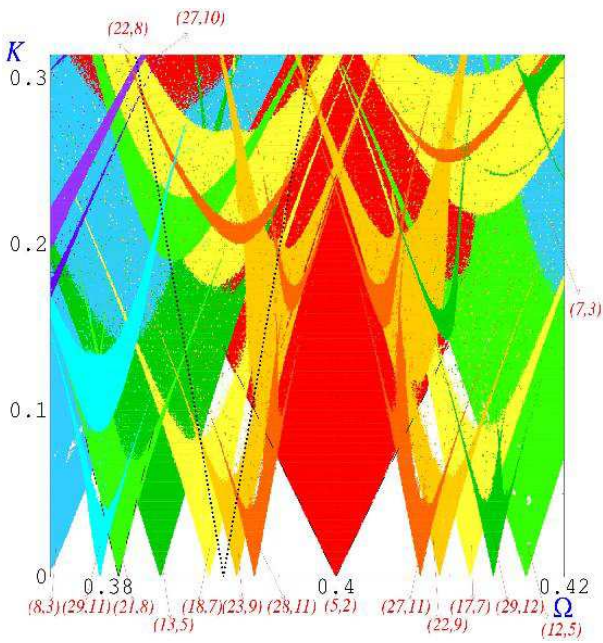}
\caption{Arnol'd tongues structure. Different colour identify different
tongues associated with different quantum accelerator modes.
The dotted line marks the
locus of experimentally explored points when $T\sim T_{2}$;
dashed lines bound regions for which $(21,8)$ and
$(5,2)$ stable periodic orbits exist, as specified by
$|K|>2\pi\sqrt{\mathfrak{p}} \left| \Omega -
\mathfrak{m}/\mathfrak{p} \right|$.} \label{Fig:tongues}
\end{figure}

In Fig.\ \ref{Fig:tongues} we plot a ``phase diagram'' to
represent the regions (tongues) where stable periodic orbits with
different values of $(\mathfrak{p},\mathfrak{m})$ are numerically
observed in the $(\Omega,K)$ parameter plane. This plot contains
the parameter space explored experimentally \cite{schlunk2003b}
using values of $T$ in the vicinity of $T_{2}=133.3 \mu$s. Close
to $K=0$, each of the stable periodic orbit regions is
wedge-shaped, with its vertex at
$\Omega=\mathfrak{m}/\mathfrak{p}$, ${K}=0$. Moving to higher $K$
inside a tongue, the periodic orbit eventually turns unstable. A
sequence of bifurcations follows, which breaks the tongue into
fragments. Fragments of different tongues intertwine and overlap
in complicated ways. A tongue may be overlapped by others even
before breaking, and such overlaps persist even at quite small
values of $K$.

Using a canonical perturbation theory
\cite{LL92,chirikov79,GRFinprp} to determine an existence
condition for a stable periodic orbit for Eq.\
(\ref{Eq:genericmap}) with a given $(\mathfrak{p},\mathfrak{m})$,
one obtains, $|K|>2\pi\sqrt{\mathfrak{p}} \left| \Omega -
\mathfrak{m}/\mathfrak{p} \right|$. In the form of an equality,
this equation accurately bounds the wedge-shaped
$(\mathfrak{p},\mathfrak{m})$ tongue near its vertex. This is
shown for the $(21,8)$ and $(5,2)$ periodic orbits by dashed lines
in Fig.\ \ref{Fig:tongues}. Numerical computation and scaling
considerations reveal the ``critical region'' where a tongue
breaks to be roughly located at kicking strengths ${K}\sim 2\pi
\mathfrak{p}^{-3/2}$ \cite{GRFinprp}. Thus, the higher the period
of an orbit, the narrower the corresponding tongue, and the lower
the ``critical value'' of $K$ at which the tongue begins to break.

\begin{figure}[htbp]
\centering
\includegraphics[width=8cm]{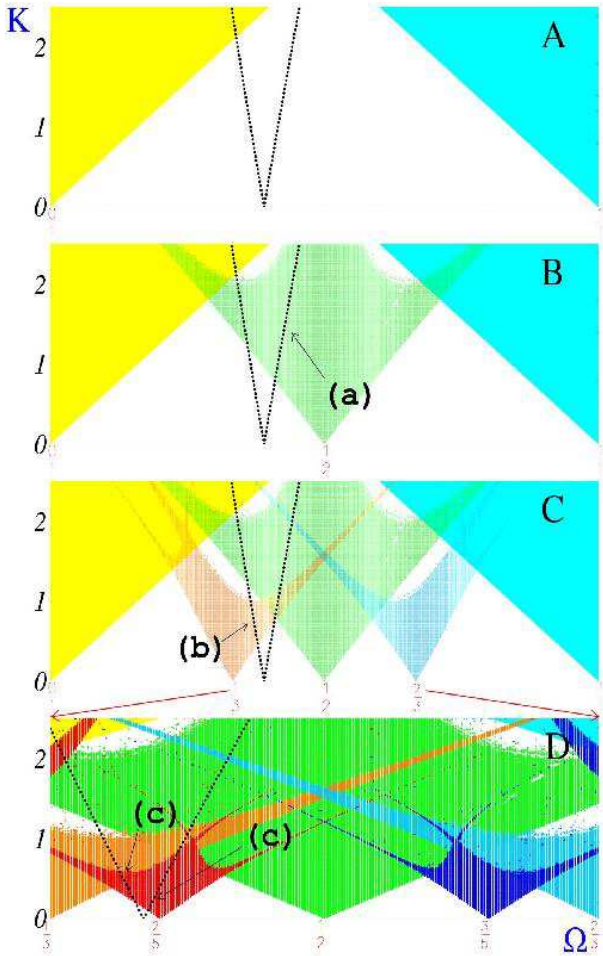}
\caption{Operation of the Farey recursion, moving from frames A to
D, for determining the experimentally observed $(2,1)$, $(3,1)$,
and $(5,2)$ quantum accelerator modes. The dotted line indicates
the locus of the experimentally explored points, while the labels
(a), (b) and (c) mark the regions in which accelerator modes are
observed. The labels correspond to those used in Fig.\
\ref{Fig:highorder}. } \label{Fig:simpletongues}
\end{figure}

The parameters corresponding to a specific experiment determine a
point in the phase diagram. If this point is inside a tongue then
a QAM may be observed. At fixed pulse number $n$, Eq.\
(\ref{Eq:acceleration}) defines a curve of enhanced population in
the $(T,p)$ plot (seen in the data of \cite{schlunk2003b}
presented in Fig.\ \ref{Fig:highorder}), due to the presence of
the $(\mathfrak{p},\mathfrak{m})$ QAM. We explore the phase
diagram of Fig.\thinspace \ref{Fig:tongues}, keeping both
$\phi_{d}$ and $n$ constant, while varying $T$. This procedure
varies $\epsilon$, $K$ and $\Omega$. The results of such an
experiment are shown in Fig.\ \ref{Fig:highorder}.  The locus of
the experimentally explored points in the phase diagram is a curve
shown by the dotted line in Figs.\ \ref{Fig:tongues} and
\ref{Fig:simpletongues}. This curve hits the $K=0$ axis at
$\Omega'=gGT_{\ell}^{2}/2\pi$, the value of $\Omega$ corresponding
to the exactly-resonant value of the kicking period $T=T_{\ell}$.
This is $0.3902$ when $\ell=2$.

Perhaps most remarkably, the values of $\mathfrak{m}$ and
$\mathfrak{p}$ corresponding to experimentally observable QAM are
determined by the Farey hierarchy of rational numbers
\cite{farey}. This representation of rational numbers is a generic
feature of mode-locking phenomena normally observed in systems
with dissipation. In this hierarchy all rational numbers in
$[0,1]$ are constructed as follows: Start from the pair
$\left(\frac{0}{1},\frac{1}{1}\right)$. At the second level the
fraction $\frac{1}{2}= \frac{0+1}{1+1}$ is introduced so that the
series consists of $\left(\frac{0}{1},
\frac{1}{2},\frac{1}{1}\right)$. On the next level the fractions
$\frac{1}{3}=\frac{0+1}{1+2}$ and $\frac{2}{3}=\frac{1+1}{2+1}$
are added. This process is continued so that if
$r_1=\frac{\mathfrak{m}_1}{\mathfrak{p}_1}$ and
$r_2=\frac{\mathfrak{m}_2}{\mathfrak{p}_2}$ are adjacent
irreducible fractions at some level, the first rational to be
added between them at the next level is their Farey mediant
$r=\frac{\mathfrak{m}_1+\mathfrak{m}_2}{\mathfrak{p}_1+\mathfrak{p}_2}$.
At no level can a rational with a denominator smaller than
$\mathfrak{p}_1+\mathfrak{p}_2$ be found between $r_1$ and $r_2$.
At each level the interval $[0,1]$ is thus divided by the Farey
fractions into Farey subintervals. As the experimental line
approaches $\Omega'$ in Fig.\ \ref{Fig:simpletongues}, it
successively intersects tongues specified by values of
$(\mathfrak{p},\mathfrak{m})$; these values determine the observed
QAM. The ratios $\mathfrak{m}/\mathfrak{p}$ are increasingly close
approximations to $\Omega'$.

To determine the tongues (and hence QAMs) appearing in the
experiment, we start from orbits with small $\mathfrak{p}$. In
Fig.\ \ref{Fig:simpletongues}A the $(1,0)$ and $(1,1)$ tongues are
presented (their vertices, at $K=0$, are outside the boundaries of
Fig.\ \ref{Fig:tongues}). These correspond to the first numbers in
the Farey hierarchy. The dotted line marking the experimental
points intersects (within the boundaries of the figure) the
$(1,0)$ tongue. In the region of intersection the stable orbit
$(1,0)$ is found. The corresponding QAMs exhibit rapid
acceleration, and at $n=30$ pulses they move beyond the
experimental window shown in Fig.\ \ref{Fig:highorder}. For
higher-order QAMs, higher orders of the Farey hierarchy are
required. At the second level the $(2,1)$ tongue, shown in Fig.\
\ref{Fig:simpletongues}B, is introduced, and $\Omega'$ is in the
interval $[\frac{0}{1},\frac{1}{2}]$. The experimental line
intersects the $(2,1)$ tongue, so a $(2,1)$ orbit (and QAM) is
found. Corresponding points are marked by (a) in Figs.\
\ref{Fig:highorder} and \ref{Fig:simpletongues}B. The third level
(Fig.\ \ref{Fig:simpletongues}C) introduces the $(3,1)$ and
$(3,2)$ tongues. The experimental line intersects both these
tongues, yet only the $(3,1)$ QAM is observed (intersection
region marked by (b) in Figs.\ \ref{Fig:highorder} and
\ref{Fig:simpletongues}C). This is because $\frac{2}{3}$ is
further than $\frac{1}{3}$ from $\Omega'$, and so the intersection
with the $(3,2)$ tongue takes place in a region where $K>2\pi
\mathfrak{p}^{-3/2}$. There are only narrow remnants of the
tongue, and the corresponding stable island is too small for a QAM
to be observable. The relevant Farey subinterval is now
$\frac{1}{3}$ to $\frac{1}{2}$. The construction can be continued
in similar fashion. In Fig.\ \ref{Fig:simpletongues}D the $(5,2)$
and $(5,3)$ tongues are introduced. Since both lines have a large
overlap with the $(5,2)$ tongue (regions marked with (c)), the
corresponding QAM appears on both sides of the resonance.
Proceeding would ideally produce all the Farey subintervals in
which $\Omega'$ belongs. Faint traces of QAMs that lie outside
this recursion may also be detected, e.g.\thinspace the white
curve (d) in Fig.\ \ref{Fig:highorder} corresponds to a $(7,3)$
mode. Note how in the experimental data it is disfavored in
comparison with the $(8,3)$ mode (curve (e)), as $\frac{3}{8}$ is
closer than $\frac{3}{7}$ to $\Omega'$.

This construction demonstrates how the Farey tree classifies the
complex structure of overlapping tongues according to those that
are most important for the description of QAMs observed for a
specific value of $\Omega'$. Furthermore, as $K \rightarrow 0$,
the value of $\frac{\mathfrak{m}}{\mathfrak{p}}$ for the QAMs seen
in the experiment converges to $\Omega'$. As $\Omega'$ is
determined by the local value of gravity, we obtain systematically
improving rational approximants of $g$. The underlying {\em
classical} mode-locking mechanism thus renders {\em quantum}
accelerator modes a robust tool for efficient quantum state
control, deep in the quantum realm.

We acknowledge support from the Royal Society,
the Lindemann Trust, the US-Israel BSF, the Minerva Centre of
Nonlinear Physics of Complex Systems, NASA, the Clarendon Bursary,
the UK EPSRC, the Israel Science Foundation, and the EU TMR ``Cold
Quantum Gases'' Network.


\begin{thebibliography}{99}
\bibitem{NoteRecoil}
$\hbar G$ is therefore equal to two photon recoils.
\end{thebibliography}
\end{document}